# Ultrafast transmission electron microscopy using a laser-driven field emitter: femtosecond resolution with a high coherence electron beam

Armin Feist, Nora Bach, Nara Rubiano da Silva, Thomas Danz, Marcel Möller, Katharina E. Priebe, Till Domröse, J. Gregor Gatzmann, Stefan Rost, Jakob Schauss, Stefanie Strauch, Reiner Bormann, Murat Sivis, Sascha Schäfer [+], Claus Ropers [*]

*4th Physical Institute - Solids and Nanostructures, University of Göttingen, Göttingen, Germany*

## Abstract
We present the development of the first ultrafast transmission electron microscope (UTEM) driven by localized photoemission from a field emitter cathode. We describe the implementation of the instrument, the photoemitter concept and the quantitative electron beam parameters achieved. Establishing a new source for ultrafast TEM, the Göttingen UTEM employs nano-localized linear photoemission from a Schottky emitter, which enables operation with freely tunable temporal structure, from continuous wave to femtosecond pulsed mode. Using this emission mechanism, we achieve record pulse properties in ultrafast electron microscopy of 9 Å focused beam diameter, 200 fs pulse duration and 0.6 eV energy width. We illustrate the possibility to conduct ultrafast imaging, diffraction, holography and spectroscopy with this instrument and also discuss opportunities to harness quantum coherent interactions between intense laser fields and free electron beams.

## Keywords
ultrafast transmission electron microscopy (UTEM), nanoscale photoemitters, nanoscale structural dynamics, ultrafast dynamics, coherent ultrafast electron pulses

## Highlights
- first implementation of an ultrafast TEM employing a nanoscale photocathode
- localized single photon-photoemission from nanoscopic field emitter yields low emittance ultrashort electron pulses
- electron pulses focused down to ~9 Å, with a duration of 200 fs and an energy width of 0.6 eV are demonstrated
- quantitative characterization of electron gun emittance and brightness enables optimized operation conditions for various applications
- a range of applications of high coherence ultrashort electron pulses is shown

## 1. Introduction
The continuing advancement of electron microscopy within physics and chemistry, materials science, and structural biology [1–3] provides us with ever-increasing precision in viewing structure and composition on the nanoscale. A detailed microscopic understanding of the structural, electronic and magnetic properties of natural and synthetic materials demands - besides atomic-scale spatial characterization - the investigation of the response of these systems to external perturbation. The growing importance of in-situ approaches in transmission electron microscopy [4], scanning electron microscopy [5], x-ray diffraction [6], scanning tunneling and atomic force microscopy [7], and other areas testify to this development.

Time-resolved experiments, following the dynamical response of a system to a pulsed excitation, represent an especially powerful form of in-situ probing, which yield direct time-

E-mail: *claus.ropers@uni-goettingen.de , +schaefer@ph4.physik.uni-goettingen.de

domain access to the character and strengths of the couplings between structural, electronic and spin degrees of freedom. Ultrafast electron [8–13] and x-ray [14–17] diffraction are well-established techniques to track structural relaxation with femtosecond temporal resolution, widely applied to homogeneous and thin film systems. The observation of spatiotemporal relaxation processes in heterogeneous systems [18–23], however, such as excitation and energy transfer across functional interfaces, is particularly challenging, requiring simultaneous nanoscale spatial and ultrafast temporal resolutions. To this end, various experimental approaches are pursued very actively at present, including time-resolved variants of scanning tunneling microscopy (STM) [24–26] and scanning near-field optical microscopy (SNOM) [27–29]. Furthermore, imaging techniques using ultrashort electron pulses such as compact point-projection electron imaging [30–33] and ultrafast scanning electron microscopy [34–36] were recently developed.

Beyond these approaches, ultrafast transmission electron microscopy (UTEM) promises to become one of the most powerful experimental tools for the investigation of ultrafast dynamics on the nanoscale, joining femtosecond temporal resolution with the vast opportunities in imaging, diffraction and spectroscopy provided by state-of-the-art electron optics. Early pioneering works at the Technical University Berlin [37], Caltech [38] and Lawrence Livermore National Labs [39] demonstrated the feasibility of pump-probe studies in electron microscopy, either in a stroboscopic fashion [40] or using single-shot imaging [41]. Motivated by various notable individual results highlighting its broad potential, time-resolved electron microscopy is currently explored in a growing number of laboratories worldwide [42–47].

Being considered one of the most exciting frontiers in electron microscopy, the area of ultrafast transmission electron microscopy is presently at a pivotal moment of its development. Facing great challenges in obtaining intense high-quality electron pulses, time-resolved electron microscopy is in particular need of benchmarking the currently achievable spatio-temporal resolution limits and electron beam figures-of-merit. Quantitative characterizations will be required to facilitate systematic progress and to connect this emerging field to the well-established, powerful experimental and theoretical framework of electron microscopy [48,49]. In order to harness the full imaging and spectroscopy capabilities of today's electron microscopes, it is highly desirable to integrate higher-brightness pulsed electron guns into the electron optics environment of a transmission electron microscope, in particular based on laser-triggered field emitter concepts. Not unlike the scientific leaps that are associated with technological breakthroughs in bright continuous electron sources in the past [49–51], significant advances in pulsed electron source quality promise a path to uncharted territory in ultrafast nanoscale dynamics.

In this contribution, we describe the first implementation of an ultrafast transmission electron microscope based on laser-triggered electron emission from a nanoscale photocathode (Ch. 2). We provide a quantitative characterization of the spatial and temporal electron beam properties for a variety of electron-optic illumination conditions (Ch. 3), demonstrating electron pulse durations down to 200 fs, energy widths of 0.6 eV and a focusability of the photoelectron beam to sub-nm dimensions. We illustrate a range of possible applications for this instrument, which include bright- and dark-field imaging, convergent beam electron diffraction (CBED) from nanoscale areas, phase-contrast imaging and Lorentz microscopy, holography and spatially-resolved electron spectroscopy (Ch. 4). Beyond adding femtosecond temporal resolution to this set of conventional electron microscopy techniques, the advanced electron beam properties of the field-emitter UTEM render it ideally suited to be applied in contrast mechanisms and phenomena that are exclusive to ultrafast electron microscopy, such as photon-induced near-field electron microscopy (PINEM) or the quantum coherent manipulation of free electron beams (Ch. 5).

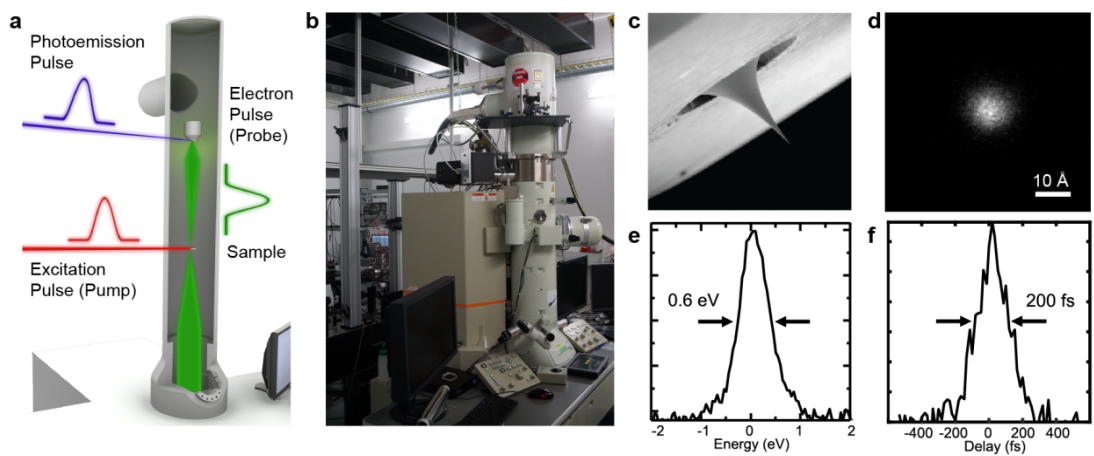

**Figure 1. Schematic setup and electron pulse properties of the Göttingen UTEM instrument.** A laser-driven Schottky field emission electron gun (a) is combined with the column of a JEOL JEM-2100F (b). Side illumination of a nanoscopic ZrO/W(100) tip emitter (c) enables the generation of ultrashort electron bunches, which can be focused down to 0.89 nm (d), with an energy width of 0.6 eV (e) and a duration of 200 fs (f) (apertured beam, at 200 kV acceleration voltage).

## 2. Instrumentation

Ultrafast transmission electron microscopy is a stroboscopic imaging technique, in which dynamics in an investigated sample are triggered by short (typically optical) excitation pulses. At well-defined delay times after excitation, the evolving state of the sample is probed by an ultrashort electron pulse (Fig. 1a). Accumulating, for a given delay time, the signal derived from many electron pulses yields a stroboscopic snapshot of the transient state of the system [8,38]. Importantly, the temporal resolution of such a pump-probe approach is given by the electron pulse duration and is not limited by the speed of the electron detector. In the past, employing photoemission driven by ultrashort laser pulses has enabled the generation of electron bunches with femtosecond duration, which are now finding increasing use in time-resolved electron imaging, diffraction and spectroscopy techniques [8,10–12,20,38,42,52–54].

The Göttingen UTEM instrument is based on a JEOL JEM-2100F transmission electron microscope, which was modified to allow for both optical sample excitation and pulsed electron emission. In contrast to previous implementations of time-resolved TEM, we employ a laser-driven Schottky emitter, which confines the photoemission to the nanoscale front facet of a ZrO/W emitter tip [35,55]. The emitter is side-illuminated with 400 nm laser radiation, focused to a spot diameter of about 20 µm (full-width-at-half-maximum (FWHM)). Optical access to the emitter tip is given through a side window on the TEM gun and by a further optical steering assembly inside the ultrahigh vacuum chamber of the electron source. For time-resolved experiments, frequency doubled femtosecond optical pulses from a regenerative Ti:Sapphire amplifier (Coherent RegA) are used at pulse energies of about 10 nJ and at a tunable repetition rate of up to 800 kHz. For alignment of the photoelectron beam into the TEM electron optics, and for characterization of electron beam properties in the space-charge-free regime, a continuous diode laser is employed at an average optical power of typically 20 mW. Utilizing the usual field geometry of a continuous Schottky source, the laser-triggered emitter is placed into an electrostatic suppressor-extractor electrode assembly, characterized by the dimensionless parameter, which allows for tuning the extraction field at the emitter apex and the divergence of the photoelectron beam. (For further details on tailoring photoelectron beams in a Schottky emitter assembly, see Ref. [56]). Electrons far from the optical-axis are cut by a hard aperture, which is placed in the electrostatic gun lens. By changing the voltages applied to these three electrodes, the electron gun can be operated in different modes, e.g., optimized for a high electron yield or a high beam coherence (cf. Ch. 3.2). Finally, after acceleration up to 200 keV, the probing electron beam is formed by the condenser system of the TEM column.

For optical sample excitation, we devised two excitation geometries by inserting mirror assemblies into the TEM column. First, access through a port conventionally used for adding an energy-dispersive X-ray spectrometer, allows for optical excitation at an angle of incidence of 55° relative to the electron beam. Second, excitation close-to-parallel to the electron beam is provided by an illumination through the pole-piece. For both cases, typical optical focal spot sizes on the sample at a central wavelength of 800 nm are about 50 µm FWHM. The delay time of optical excitation, optical fluence, polarization state and focal spot position can be changed in an automatized fashion, allowing for versatile means to investigate the dependence of sample dynamics on the optical excitation.

## 3. Implementation of a laser-triggered field emitter source in the UTEM

### 3.1 Localized Photoemission from needle-shaped photocathodes

Over the last decade, various approaches have been undertaken to enhance the beam quality of ultrafast electron sources for time-resolved experiments, including a tailoring of photoemission laser wavelength [57], photocathode workfunctions and materials [58–60], as well as photoemission spot sizes [61,62], and by enhanced extraction fields in radio-frequency cavities [10] and at sharp emitter needles [20]. Also alternative approaches using cold atomic gases are pursued [63–65].

The nanoscale localization of the emission area in needle-shaped laser-driven photocathodes promises particularly coherent electron beams and has thus been intensely studied, recently [66–78]. Compared to beams derived from state-of-the-art planar photocathodes [8,13,57,58,61,79,80], electron pulses emitted from tip emitters occupy a significantly reduced area of transverse phase space (cf. Fig 2). In particular, providing the same advantages as static tip field-emitters for conventional electron microscopy [48,50], ultrafast nanoscale electron sources are crucial for time-resolved imaging applications, which require either a sharply focused or a well-collimated, highly coherent electron beam (cf. Fig 6, Chap. 4). Consequently, they have enabled ultrafast low energy electron diffraction [20], ultrafast scanning electron microscopy [35,36] and fs-point projection imaging [30–33,81,82], and, as we demonstrated recently, ultrafast transmission electron microscopy with highly coherent electron beams [42].

In tip-based ultrafast photoemission, localization is typically achieved by employing optical field enhancement at the apex in combination with a high nonlinearity of the photoemission process [32,33,66–68,70,71,83]. However, processes such as higher-order multi-photon and strong-field photoemission exhibit significantly broadened photoelectron energy distributions [70,83], which limits their use for ultrafast electron imaging and spectroscopy applications. In a regime of lower nonlinearity, sufficient beam qualities can be achieved by partially localized photoemission at the emitter apex in combination with a small electron energy bandwidth, as recently demonstrated using two-photon photoemission (cf. Fig 3h) from tungsten needle emitters [42,56]. In principle, a linear photoemission regime may be considered ideal, due to the low thermal load on the emitter as well as a simple tunability of the electron pulse length via the laser pulse duration, provided nanoscale localization is ensured.

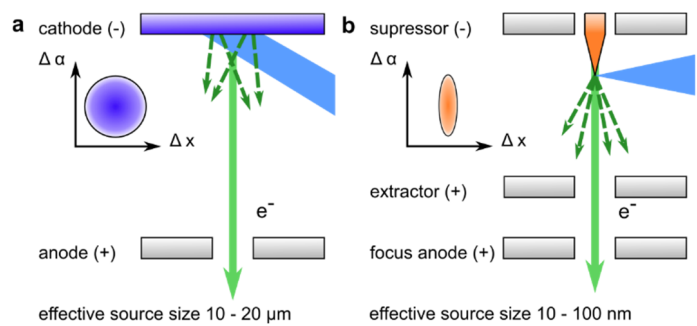

**Figure 2. Transverse beam properties of photoelectron beams emitted from laser-triggered electron sources. (a) The micrometer-scale emission area (typical diameter of 10-20 µm, governed by the laser spot diameter) of a flat photocathode results in photoelectron bunches occupying a large transverse phase space volume. (b) Localized electron emission from the apex of a nanoscopic tip (typical apex diameter of 10-100 nm) allows for a minimization of the phase space volume occupied by the photoelectrons, resulting in significantly enhanced beam coherence. Insets: Sketch of the transverse phase space distributions for the respective source geometries (Δx: spatial coordinate perpendicular to beam propagation axis, Δα: angular coordinate relative to the optical axis).**

In this work, we introduce a linear photoemission process for UTEM, in which photoemission is localized by chemically tuning the work function of the emitter front facet. Specifically, we make use of single-photon photoemission from zirconium oxide covered (100)-oriented single crystalline tungsten tips (ZrO/W(100), cf. Fig 3a-d) [35,55] and quantitatively characterize their performance in UTEM.

Schottky field-emitters based on ZrO/W(100) without laser excitation are routinely used and well-characterized as continuous electron sources with high brightness [84]. Operated at an elevated temperature of about 1800 K and with electric extraction fields in the range of 0.5-1 V/nm, the (100) front facet of the emitter is covered by a ZrO overlayer, which reduces the work function down to about 2.9 eV. A further lowering of the work function is achieved by the applied extraction field (Schottky effect), resulting in intense thermal electron emission from such sources [85,86] (cf. Fig 3f).

In photoemission mode, we operate the emitter at a reduced temperature (below 1400 K), so that static electron emission is fully suppressed, and photoelectrons are only generated with the laser focus placed at the tip apex. At an illuminating wavelength of 400 nm, the extracted photocurrent scales linearly with the incident laser power (Fig. 3i), signifying a single-photon photoemission process (Fig. 3g), and the strong localization of photoemission at the (100) front-facet is demonstrated by imaging the photoelectron source using the TEM optics (Fig. 3e).

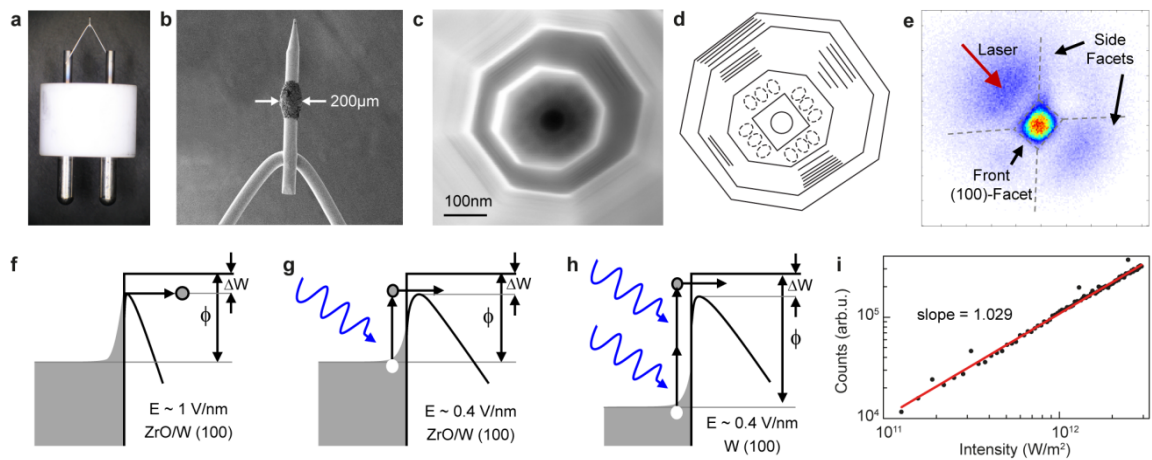

**Figure 3. Operation principle of a laser driven Schottky field emitter and its geometry. (a,b) Single-crystalline, (100)-oriented tungsten tip emitter attached to a heating filament. The ZrO$_2$ reservoir deposited on the emitter shank is visible in (b). (c,d). Top-view SEM micrograph (c) and corresponding schematic (d) of the faceted apex shape after high-temperature tip conditioning. (e) Photoemission pattern demonstrating the localized photoemission from the (100) front facet (central intense spot) with minor contributions from (100)-equivalent crystal surfaces at the emitter shaft (weak intensity side-lobes). (f-h) Schematic energy diagrams at a Schottky-lowered potential barrier for continuous thermal electron emission (f), and linear (g) and two-photon (h) photoemission. (i) Photoemission current scales linearly with incident laser power density, as expected for a single-photon photoemission process.**

The photoemitter can be stably operated for an extended period of time (>48 h) at high photoelectron currents. Over time, a slow decrease in photoemission efficiency is observed, which, however, can be fully reversed by flashing the tip to temperatures above 1700 K. Further utilizing these emitters in electron microscopy requires a detailed analysis of their performance in an electrostatic lens assembly [56,71] and of the resulting beam properties in the TEM column, which we will address in the following.

## 3.2 Characterization of spatial Beam Properties using photoelectrons

The quality of an electron beam, or radiation source in general, for imaging, local probing or diffraction is commonly assessed by its normalized brightness $B_n$, here given as the amplitude of a Gaussian-shaped beam [87]:

$$B_n = \frac{I}{4\pi^2 \varepsilon_{n,rms}^2},$$

where I denotes the beam current. The quantity $\varepsilon_{n,rms}$ is the so-called normalized r.m.s. emittance [88], which describes the transverse phase space area occupied by the particle beam, and accounts for its transverse coherence properties. For aberration-free electron optics, the normalized beam brightness is a conserved quantity, unaffected by apertures and imaging lenses. Lens-aberrations and space-charge interactions within the (pulsed) electron beam result in a decrease of beam brightness [88,89]. In order to quantitatively characterize the brightness of the UTEM photoelectron beam for different source and condenser settings, we measure the caustic of the focused beam in the sample plane. The electron beam current is recorded with a calibrated CCD camera, and the emittance is derived from the width of the electron focal spot and its convergence angle.

Specifically, for a focused beam, the normalized r.m.s. emittance along one transverse direction (x) is given by [88]:

$$\varepsilon_{n,rms,x} = \beta\gamma \cdot \sigma_x \sigma_{\alpha x},$$

with $\beta = v/c$ (v: electron velocity, c: light velocity) and the Lorentz factor. Here, the width of the focused electron beam along the x-axis and its corresponding distribution in propagation angles are characterized by the standard deviations $\sigma_x$ and $\sigma_{\alpha x}$, respectively. The obtained focal spots are largely symmetric along the two transverse directions (x and y). Small asymmetries are accounted for by defining the overall emittance $\varepsilon_{n,rms}$ as the geometrical mean value of the emittance along the x- and y-directions.

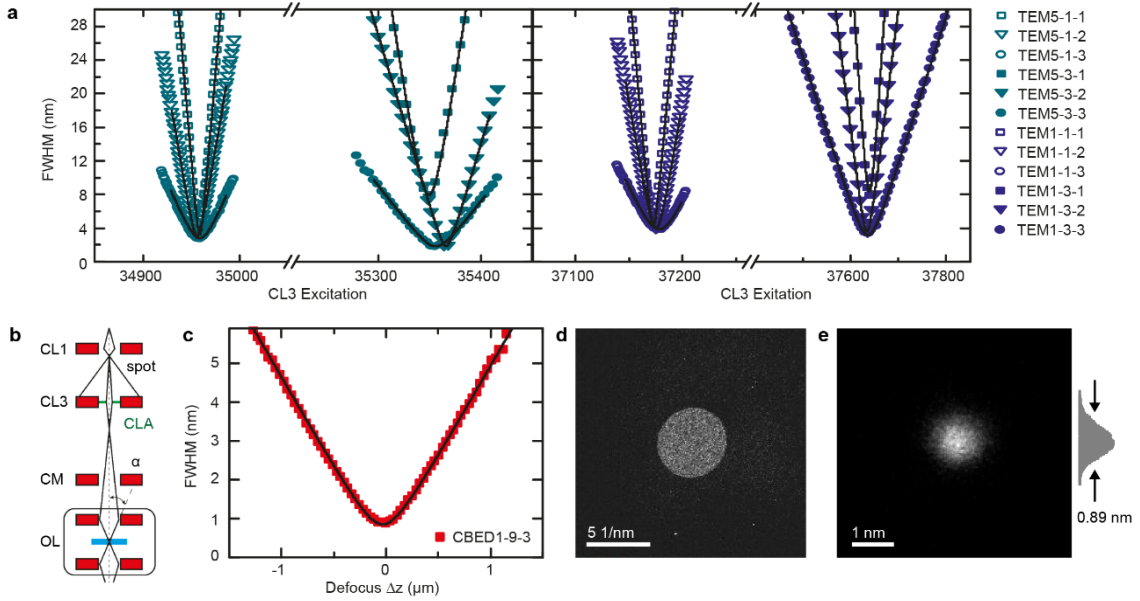

**Figure 4. Characterization of the transverse electron beam properties. (a) Beam caustics for a range of condenser settings 'TEM a-b-c', with transverse beam coherence 'a' (spot size), convergence angle 'b' (alpha) and condenser aperture 'c' (CLA) in TEM illumination mode. (b) Illustration of the TEM column illumination system. The beam coherence is adjusted by the first condenser lens (CL1), the convergence angle is set by the condenser mini lens (CM) and the beam is limited by the condenser aperture (CLA). The second condenser lens (CL3) is adapted accordingly to form a focus at the sample plane. (c) Beam diameter in the focal plane of the imaging system as a function of the second condenser lens excitation CL3. (d,e) Angular distribution and beam profile at the minimum focal spot. ((b-d): convergent beam illumination mode (CBED)).**

At different condenser settings, the spatial beam size in the sample plane is shown in Figure 4a, varying the excitation strength of the second magnetic lens in the condenser system (CL3) and thus placing the electron focus at different positions relative to the sample plane [90,91]. An effective beam caustic, i.e. the beam envelope as a function of the position along the electron optical axis (Fig. 4c), is then extracted by taking into account the beam convergence angle measured in diffraction mode. The exemplary caustic shown in Fig. 4c exhibits a minimum focal spot size $d_{\min}$ of only 0.89 nm at a convergence semi-angle α of 6.7 mrad (cf. Figs. 4d,e). For this setting, we obtain an r.m.s.-emittance of the photoelectron beam of only 1.71 nm*mrad - a value which is only one order of magnitude larger than the minimum emittance given by the uncertainty principle [92], which is easily derived to be $\varepsilon_{n,rms} = \hbar / (2 \cdot c \cdot m_0)$ - and a temporally averaged, normalized brightness of $2.85 \cdot 10^7$ A/(m²sr).

Considering the transverse coherence length [93] given by

$$\xi_{c,x} = \frac{\hbar}{m_e c} \cdot \frac{\sigma_x}{\varepsilon_{n,rms,x}},$$

such a beam readily allows for 1-µm-scale coherence lengths (cf. Fig. 7g) when spread to a diameter of about ten micrometers (FWHM). The degree-of-coherence $K = \xi_{c,x}/(2\sigma_x)$ [94] is 11.2%, which, equivalently, can be described by a beam quality factor of M²=1/K=8.9.

In UTEM, the beam current, emittance, and degree-of-coherence can be precisely tailored, depending on the relative potentials applied to the suppressor and extractor electrode in the electron source, and the settings of the condenser systems (selected conditions shown in Tab. 1). In particular, for an extractor-suppressor field geometry (Γ=1.11, see Ref. [56]) allowing for a high transmission through the beam-limiting aperture in the source, a beam current of about

|  | TEM Setting | | | $d_{\min}$ | α | I | $\varepsilon_{n,rms}$ | K | $B_n$ | $B_{np}$ |
|---|---|---|---|---|---|---|---|---|---|---|
|  | Spot | α | CLA | nm | mrad | fA | nm mrad | % | $10^6$ A/m²sr | $10^{12}$ A/m²sr |
| TEM Mode, 120 kV | 1 | 3 | 1 | 27.7 | 25.4 | 459 | 153 * | 0.13 * | 0.50 * | 0.62 * |
| gun setting A | 1 | 3 | 2 | 16.3 | 13.3 | 107 | 47.2 * | 0.40 * | 1.22 * | 1.52 * |
| (high transmission) | 1 | 3 | 3 | 10.7 | 5.3 | 18.2 | 12.4 | 1.56 | 3.01 | 3.76 |
|  | 5 | 3 | 3 | 6.23 | 5.5 | 2.74 | 7.40 | 2.61 | 1.27 | 1.59 |
|  | 1 | 3 | 4 | 9.37 | 2.1 | 1.47 | 4.24 | 4.55 | 2.08 | 2.60 |
| TEM Mode, 200 kV | 1 | 1 | 1 | 4.40 | 10.6 | 101 | 13.5 | 1.43 | 13.95 | 17.47 |
| gun setting B | 3 | 3 | 2 | 2.33 | 9.6 | 18.1 | 6.47 | 2.98 | 10.93 | 13.69 |
| (high coherence) | 5 | 1 | 3 | 2.65 | 2.5 | 0.97 | 1.90 | 10.1 | 6.76 | 8.47 |
| CBD Mode, 200 kV gun setting B | 1 | 9 | 3 | 0.89 | 6.7 | 0.30 | 1.73 | 11.2 | 2.85 | 3.18 |

**Table 1. Electron beam properties for a range of TEM illumination conditions, with the electron gun operated at a high transmission of the electron emitter (gun setting A) and high coherence mode (gun setting B). The beam properties at the sample are characterized for a specific condenser setting (spot, alpha, CLA) by the minimum focal spot size (FWHM), semi-convergence-angle, overall electron flux (continuous laser illumination, 20 mW), degree of coherence and averaged normalized brightness. A normalized peak brightness is obtained by scaling to typical pulsed beam conditions for space-charge free operation (300 fs, 2 mW, 250 kHz). All beam properties are measured at the sample position and, therefore, contain both contributions from the intrinsic source properties as well as aberrations from the imaging optics, including the spherical aberration of the objective lens at high convergence angles (indicated by *).**

460 fA is obtained with a focal spot size of 28 nm, whereas in a high-coherence mode (Γ=0.55), local probing with sub-nm spot sizes is achieved.

To compare our emitter with a continuous electron source, we derive a normalized beam peak brightness $B_{np}$ by scaling the time averaged brightness with an effective duty cycle $D_{eff} = \sqrt{2} \cdot f \cdot \sigma_t$ containing the laser repetition rate $f$ and a typical full-width-at-half-maximum electron pulse duration $\tau = \sqrt{8\ln(2)} \cdot \sigma_t$ at the sample, i.e. $B_{np} = B_n / D_{eff}$. For operation in regimes not affected by space-charge, we arrive at a norm. peak brightness of $1.75 \cdot 10^{13}$ A/m²sr, which is comparable to reported time-averaged values for a conventional Schottky field emitter [85,86].

For multi-electron pulses, longitudinal and transverse space-charge broadening can be observed, although the effect on the transverse beam properties is rather moderate. In particular, even for electron pulses spectrally broadened to about 5 eV, we only observe a slight degradation of the transverse beam properties (cf. Sec. 3.3 and Fig. 5d,h). Generally, we note that space-charge effects at nanoscopic tip emitters are expected to be reduced compared to flat photocathodes, due to the high intrinsic extraction fields (~ 1V/nm) and divergent beam trajectories.

### 3.3 Characterization of temporal electron bunch properties

Electron pulse durations are experimentally determined by laser-electron cross-correlation [95–102]. In particular, as discussed in more detail in Ch. 5, electrons which traverse an intense optical near-field experience inelastic scattering, resulting in photon sidebands in the electron energy spectrum [42,95].

Figure 5 (a,e) shows electron energy spectra as a function of the delay between optical near-field excitation and electron arrival time at the sample, for two experimental conditions differing only in the number of electrons per pulse. In the first case, with minimized space-charge broadening (Fig. 5a), photon sidebands on the energy-gain and -loss side are visible only within a narrow delay window around the temporal overlap between electron and laser pulses. It should be noted that due to the convolution of the electron pulse profile with the nonlinear electron-light interaction across the 50-fs optical pulses, the temporal interval with considerable photon sideband intensity represents a reliable upper bound to the electron pulse duration [42,96,100]. Hence, the total number of gain-scattered electrons for different delay times (Fig. 5c) yields a measurement of the electron pulse duration of 300 fs.

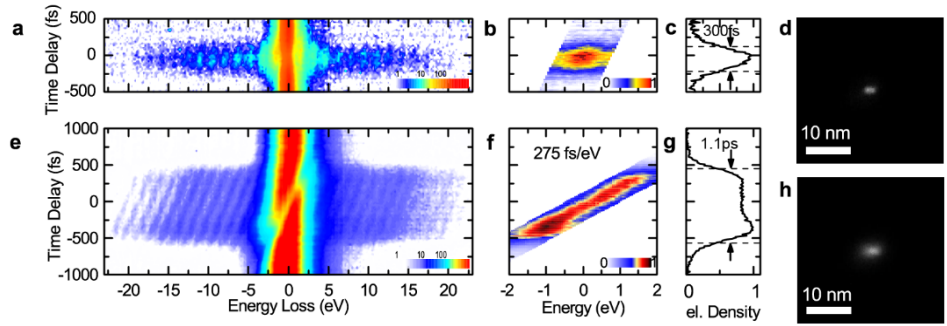

**Figure 5.** Temporal pulse characterization by electron–photon cross-correlation for electron pulses in the absence (a-c) and presence (e-g) of space-charge broadening (260-times higher pulse charge in (e-g)). (a,e) Electron energy-loss spectra as a function of time delay (logarithmic color scale). (b,f) Extracted energy- and time-resolved structure of electron pulses, revealing a linear chirp in the case of space-charge broadened electron bunches and a longer pulse duration (c,g). Comparing the focal spot sizes with (d) narrow energy distribution (here: thermal emission) and (h) electron pulses with strong spectral broadening (5.3 eV), there is only a slight degradation of the transverse beam properties (increase of minimum spot size from 2.2 to 3.5 nm). For the measurements shown the gun was operated at high coherence mode and 120 kV acceleration voltage, spectra recorded in the proximity of a nanoscopic gold tip (cf. Fig. 8b and Ref. [42]).

For electron pulses containing a larger number of electrons at emission, space-charge effects become operative, accelerating electrons at the leading edge of the bunch, and decelerating electrons at the trailing end [103]. As a result, the electron bunch width is temporally broadened, illustrated by the appearance of photon side-bands over a delay interval of about 1 ps (Fig. 5e,g) and by the increase in the energetic width of the zero-loss peak from 0.85 eV to about 3.5 eV. The associated pronounced chirp of the electron pulse is visible as an inclination of the individual side-bands in the EELS maps (Fig. 5e) [96,104]. Averaging over each sideband in the delay-energy maps gives a direct representation of the longitudinal phase space structure of the electron bunch (Fig. 5 b,f). Specifically, for the space-charge broadened pulse considered here, we extract a chirp of 275 fs/eV and a momentary energetic width of 0.65 eV (measured for a single delay value), close to the overall spectral width of the non-broadened pulse. Such a strong correlation between the longitudinal electron position (i.e. arrival time at the sample position) and electron energy, indicates that space-charge forces predominantly lead to shearing in the longitudinal phase space, approximately preserving the bunch's phase space volume. Therefore, in future UTEM implementations, phase-locked radiofrequency [105] or THz [10,106] fields may be incorporated into the TEM column to temporally and spectrally re-compress multi-electron bunches, as already successfully applied in ultrafast electron diffraction beam lines [10,80,104,107].

While for the high-charge bunches considered in single-shot imaging, space-charge induced deterioration of transverse beam properties presents a major challenge [13,39,108,109], space charge only weakly affects the few-electron pulses studied here. Specifically, the minimal focal spot size displays only a minor increase, when comparing thermal and spectrally strongly broadened photoemission, here from 2.2 to 3.5 nm with an energy width increase from 0.7 eV to 5.3 eV (cf. Fig. 5d,h). We note that the electron bunches generated by single-photon photoemission, in continuous or non space-charge broadened fs-operation, display comparable transverse beam properties to those thermally extracted from the emitter with identical electrostatic gun settings.

# 4. Selected Applications

Low-emittance photoelectron beams, as demonstrated here, are ideally suited for ultrafast transmission electron microscopy with nanometer-scale spatial resolution (Fig. 6). Depending on the parameters of the condenser electron optics, the small transverse phase space area occupied by photoelectron bunches facilitates few-nanometer-scale electron foci, and collimated beams with µm-scale transverse coherence lengths, respectively (cf. Tab. 1).

For all experiments, pre-alignment of the electron column in a continuous mode is possible by raising the temperature of the tip and entering the thermal emission regime of the Schottky emitter. Even after modification of the electron gun, high emission currents of several µA can be generated, thus enabling an in-situ characterization with high quality electron micrographs of the sample before and after time-resolved experiments, (e.g. high resolution TEM, cf. Fig. 7a). Switching between thermal and photoemission mode requires less than 1h.

Figure 7 shows a set of examples demonstrating the present imaging, diffraction and spectroscopy capabilities of the Göttingen UTEM instrument in photoemission mode. With typical acquisition times of about 5-60 s, electron micrographs are obtained which exhibit sufficient signal-to-noise ratio to map, for example, bending contours (Fig. 7b) as well as nanoscale magnetic textures in Lorentz mode (Fig. 7c). High-quality electron diffraction patterns can be recorded both with parallel (Fig. 7d,e) and convergent incident beams (Fig. 7f). We note that, despite working in the single- to few-electron per pulse regime, low intensity diffraction features arising from charge density waves can be clearly discerned. In addition, a high transverse coherence length of 1.2 µm can be determined by scattering from a mesoscopic grating structure (463 nm spacing) (Fig. 7g). This will enable ultrafast electron holography (Fig. 7h) for the measurement of time-dependent electric and magnetic fields. Finally, the narrow spectral width of the electron beam of about 0.6 eV allows for recording well-resolved electron energy loss spectra, which will facilitate the study of charge carrier dynamics and electronic structure in complex materials (Fig. 7i).

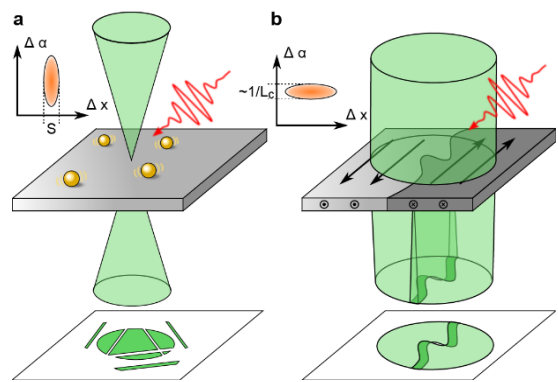

**Figure 6. Applications of low-emittance electron pulses in ultrafast electron imaging. (a) Electron pulses focused to nanoscale spot sizes allow for local ultrafast probing, including ultrafast convergent beam electron diffraction (sketched here) and ultrafast electron energy loss spectroscopy. (b) For collimated low-emittance electron pulses, µm-scale transverse coherence lengths are achievable, enabling phase-sensitive electron imaging techniques, such as ultrafast Lorentz microscopy (sketched here) as well as time-resolved variants of holographic techniques. Insets: Occupied area in the transverse phase space for a focussed and a collimated beam, respectively (S: focal spot size; $L_C$: transverse coherence length).**

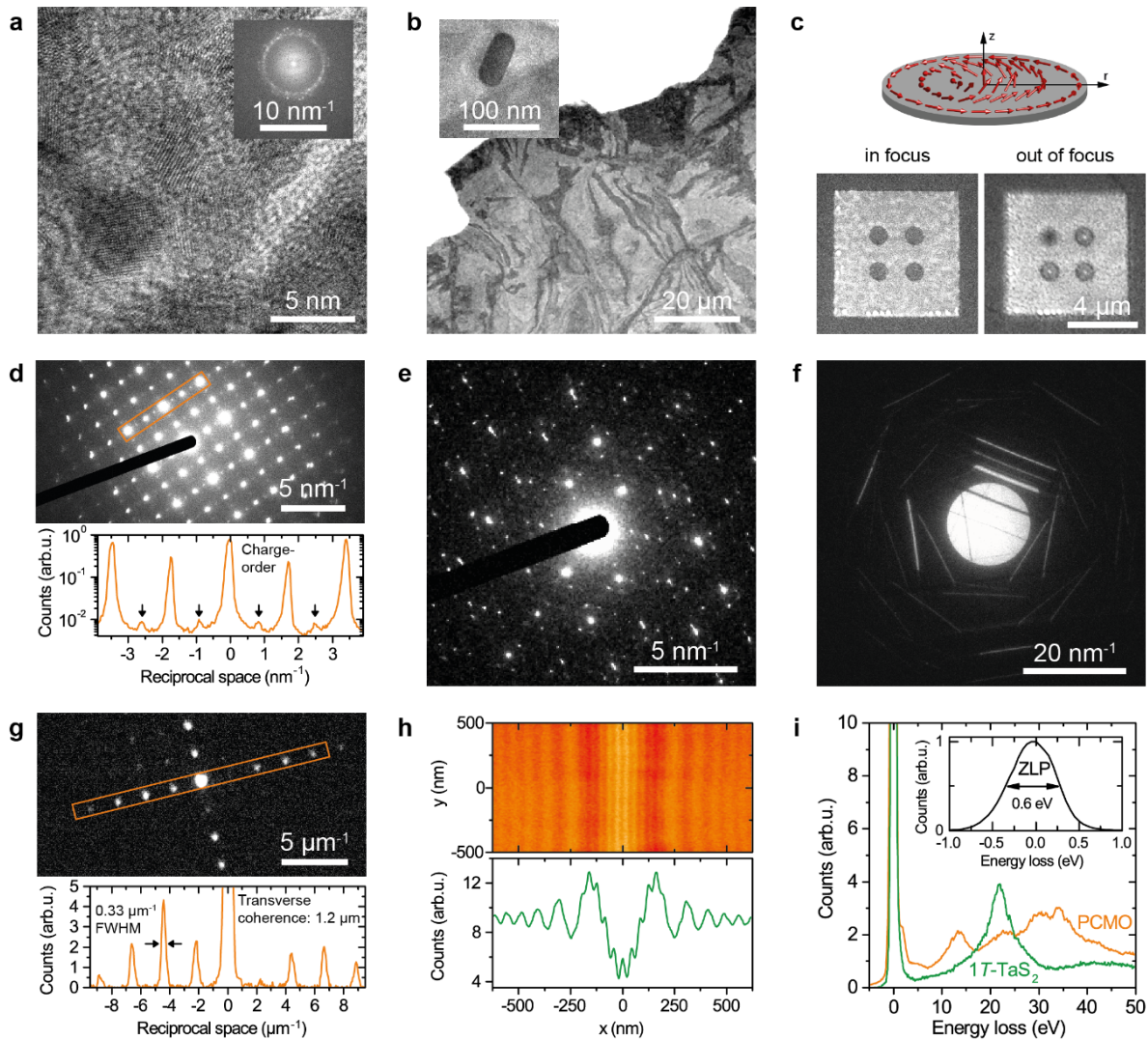

**Figure 7. Exemplary experimental results achievable with the current status of the UTEM instrument. (a)** High-resolution TEM (HRTEM) micrograph of Au/Pd particles on an amorphous carbon film. Visible lattice planes with 2-Å spacing demonstrate the resolution capabilities of the modified instrument (here: using thermal electron emission). Inset: Fourier transform of a four times larger sample region. **(b-i)** Measurements acquired with photoelectron beams (typical acquisition times 5-60 s) and at an electron energy of 120 keV. **(b)** Bright-field image of an ultra-microtomed 50 nm thin sample of 1T-TaS$_2$ showing bending contrast of the thin-film membrane. Close-up: drop-casted gold nanorod on the sample surface. **(c)** Lorentz imaging provides magnetic contrast in UTEM as demonstrated for permalloy islands on a silicon nitride support. Out-of-focus image reveals the existence of a magnetic vortex in each of the four islands (visible as black and white features, respectively, depending on vortex orientation). Magnetic structure of a single vortex is schematically depicted in the upper panel. **(d)** Diffraction pattern of the charge-ordered phase of an ion-polished PCMO (Pr$_{0.7}$Ca$_{0.3}$MnO$_3$) plan view sample. Weak superstructure spots are visible halfway between the lattice reflections. **(e)** Diffraction pattern of the nearly commensurate charge density wave (NC-CDW) phase of 1T-TaS$_2$. The first-order NC-CDW diffraction spots are hexagonally arranged around structural reflections. **(f)** Convergent beam electron diffraction (CBED) pattern of an exfoliated 100 nm thick single-crystalline graphite flake. **(g)** High dispersion diffraction pattern of a 463 nm spaced grating replica, demonstrating 1.2-μm transverse coherence lengths. **(h)** Electron hologram obtained using a Möllenstedt biprism at a filament voltage of 9V, emphasizing the photoelectron coherence properties achievable in the UTEM. **(i)** Electron energy loss (EEL) spectra of 1T-TaS$_2$ and PCMO. Inset: zero-loss peak (ZLP) with a FWHM of 0.6 eV.

# 5. Optical interactions with free electron beams in the field-emitter UTEM

Besides adding ultrafast temporal resolution to widely established electron microscopy techniques, the high coherence electron beams in UTEM also generate research directions completely outside the realm of conventional electron microscopy. A prominent class of new phenomena involve the interaction of the pulsed free-electron beam with intense optical fields [42,95,98,110–112]. The exchange of energy and momentum between electromagnetic fields and free electrons provides multiple avenues of study, which include (i) the temporal characterization of ultrashort electron pulses (see Ch. 3.3) [98,100], (ii) the nanoscale mapping of optical near-fields [95,113], (iii) the active manipulation of free-electron beams [42,114], and (iv) the study of fundamental quantum optics phenomena [42,95,115].

In inelastic electron-light scattering (Fig. 8a), a beam of free electrons passes through the optical near-field of an illuminated nanostructure. The electrons exchange energy with the optical field in integer multiples of the incident photon energy, and consequently, the interaction transforms an initially narrow kinetic energy distribution into a symmetric spectral comb composed a number of populated sidebands (see spectra in Fig. 8a). Physically, this process is closely related to electron energy loss (EEL) or cathodoluminescence (CL) at optical nanostructures [116–120]. All of these processes are facilitated by the near-field localization of optical excitations, which relax the requirement of conserving the total momentum in the electron and light fields alone by transferring excess momentum to a nanostructure. In EEL and CL, a swift electron passing a structure induces an optical polarization, with which it interacts [110]. Energy loss and cathodoluminescence then correspond to a spontaneous transition in the free-electron state, resulting in material absorption, the emission of far-field radiation or near-field excitations such as plasmons [110,121]. In essence, the inelastic

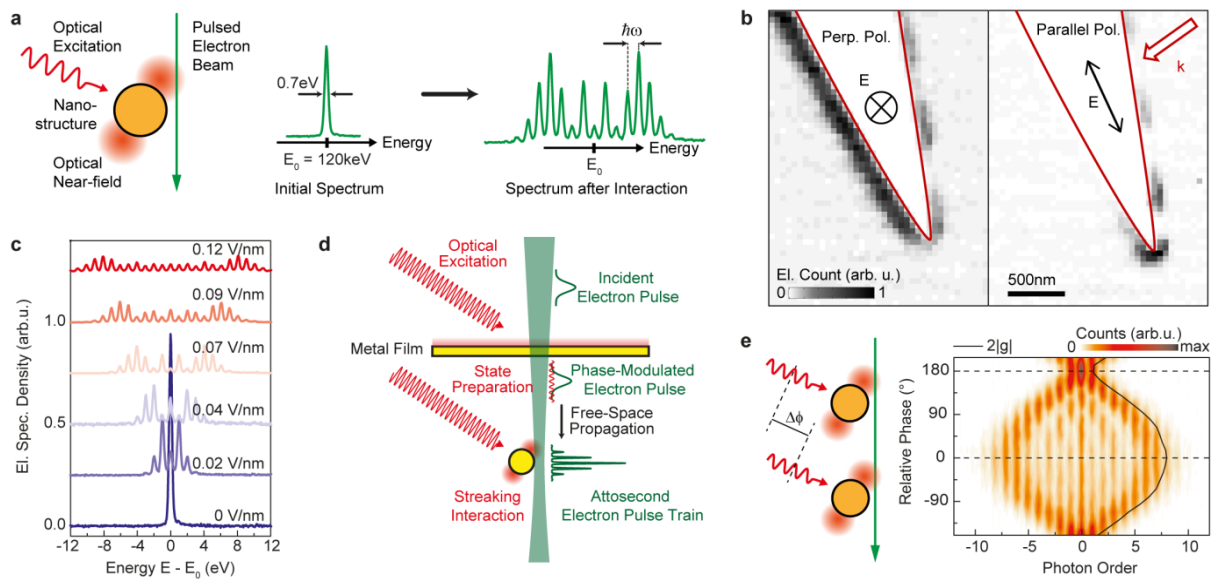

**Figure 8. Applications of quantum coherent electron light interactions in optical near-fields within an ultrafast TEM.** (a) The electron beam traversing an intense optical near-field develops into a comb of spectral sidebands, with (c) populations given by the optical field strength (assuming spatially and temporally homogeneous illumination). (b) Raster-scanning the electron beam allows for a quantitative mapping of the near-field distribution (S-PINEM) at a nanostructure (here: nanoscopic gold tip side-illuminated along the direction indicated by k, with different optical polarizations and at 800 nm wavelength). (d) Dispersive broadening of the optically phase modulated single electron wavefunction leads to the formation of an attosecond train, with a temporal spacing given by the optical period. (e) Consecutive interactions with multiple near-fields enable Ramsey-type electron light interferometry, where the first interaction can be either cancelled out or enhanced depending on relative excitation phase. (a-c) Adapted from Ref. [42] , (e) adapted from Ref. [114])

interaction described here may be viewed as the stimulated absorption and emission variants of these spontaneous processes.

This prompt interaction can be used in a variety of applications, and it was initially motivated by the desire to map optical near-fields in an approach termed photon-induced near-field electron microscopy (PINEM) [95,111,122]. In our UTEM, we implemented a scanning version of this technique (S-PINEM), in which a focused electron beam is scanned in the vicinity of an optically excited nanostructure, and an electron spectrum is recorded for every scan position (compare Ref. [42]), providing a quantitative measure of the optical near-field amplitude. Figure 8b displays the fraction of inelastically scattered electrons at a metallic nanotip illuminated with polarizations perpendicular (right) and parallel (left) to the tip axis. Different optical near-field modes and a field-enhanced region at the apex of the conical structure (parallel illumination) are clearly visible in these images.

For increasing field strength, the spectra produced from this interaction exhibit a characteristic broadening and spectral oscillations in the individual sideband amplitudes (Fig. 8c). In Ref. [42], we demonstrated that these features are caused by multilevel Rabi oscillations in the free electron states separated by the photon energy. In a spatial representation of the electron states, the interaction results in a sinusoidal phase modulation of the incident wave function [42,111,122]. As a result, dispersive propagation of the wavefunction after the interaction will cause a reshaping of the electron density subsequent to the interaction (Fig. 8d). Specifically, as shown in Ref. [42], the momentum modulation will cause a temporal focusing of the electron density into a train of attosecond pulses downstream in the electron microscope, at propagation distances in the one-to-few millimeter range (depending on the light frequency, the electron energy and the optical excitation strength). With a further nanostructure located in the region of the temporal focus (bottom in Fig. 8d), the arising attosecond pulse structure may be probed with a second, properly timed interaction. Representing a feasible means to generate attosecond electron pulses within the UTEM, this scheme will in the future allow for entirely new forms of optically phase-resolved electron microscopy and the study of electronic or structural dynamics with sub-femtosecond precision.

We recently applied the concept of multiple quantum coherent interactions with the same free-electron state in an experiment sketched in Fig. 8e (for details, see Ref. [114]). Here, two optical nanostructures are separated by several micrometers, i.e., at a distance for which no substantial electron density reshaping occurs. The electron beam sequentially interacts with two phase-locked optical near-fields, the relative phase of which can be precisely controlled. The color-coded image in Fig. 8e displays the resulting electron energy spectra for a variation of the relative phase. The final width of the energy spectrum, and thus the total impact of the interaction with the free electron beam, is a strong function of this relative phase. This observation highlights the quantum coherent nature of these sequential interactions, in that the second action either cancels out or enhances the action of the first [114]. Such phase-controlled multiple interactions may form the basis of different variants of electron interferometry or time-domain holography, and – combined with optically excited materials inserted into the interferometer gap – may yield detailed information on nanoscale dephasing mechanisms on ultrashort timescales.

More generally, the demonstration of coherent and phase-sensitive optical near-field scattering opens up an exciting research path in the active quantum manipulation of electron wave functions. In particular, any electron wave packet of sufficient longitudinal and transverse coherence will directly carry the entire spatial and temporal amplitude and phase information of that optical near-field in a holographic fashion. Governing the further evolution of the electron probability density in space and time, such optically-produced holograms create far-reaching opportunities for coherent control schemes using free electrons, including the generation of specific transverse profiles and orbital angular momentum states [123–125], or the arbitrary formation of temporal electron pulse structures.

# 6. Conclusion & Outlook

In conclusion, we described the present status of our development of ultrafast transmission electron microscopy using laser-triggered field emitters. We provided quantitative beam characterizations and presented exemplary imaging, diffraction and spectroscopy data recorded with this instrument.

Generally, nanoscopic cathodes offer superior performance over flat photocathode designs if nanoscale probing or high-sensitivity phase contrast are desired. Historically, many novel techniques in transmission electron microscopy, like sub-nm STEM probing and holography, were only enabled by the introduction of high-brightness field emitting electron sources. In the same way, the integration of tip-shaped photocathodes with an energy spread of less than 1 eV and the peak brightness of a conventional Schottky emitter opens up new frontiers in electron microscopy and the study of nanoscale dynamics.

Nonetheless, further improvements in average source brightness are highly sought after for the further proliferation of the technique, and for widening the scope of applications. Alongside further development work on the actual source, an optimized electron gun geometry could increase the overall electron transmission from tip to sample, thus decreasing space-charge effects for a given final pulse charge, while minimizing aberrations and propagation-induced temporal pulse spread. Additional strategies will extend the practically accessible set of scientific problems with UTEM, such as devising intelligent drift correction schemes to allow for longer integration times. Moreover, the integration of aberration probe correction would enhance the acceptance of the TEM column and thus increase the total achievable pulse charge for a certain target specification in spatial electron beam parameters.

The very robust free-space laser excitation of the sample, as well as the long-term stability of our short-pulsed electron source, facilitate stable measurements on the nanoscale for periods exceeding 24h. This will enable the study of structural, electronic and spin dynamics unprecedented spatial and temporal resolution. At present, ultrafast electron diffraction and spectroscopy from few-nanometer regions is well within the range of capabilities of the approach, as is phase contrast imaging and Lorentz microscopy of samples exhibiting usual phase shifts. The generation and application of attosecond electron pulse trains will ultimately allow electron microscopy to enter the realm of attosecond science [126], adding nanoscopic spatial resolution to this exciting research field. With all these prospects, we believe that field-emitter-based UTEM technology will foster a greatly enhanced understanding of spatiotemporal dynamics, energy transport and relaxation processes on the atomic scale.

# Acknowledgments

We acknowledge useful discussions and supporting interactions with Konrad Samwer, Christian Jooss, Michael Seibt, Cynthia Volkert, Markus Münzenberg, Max Gulde and Patrick Peretzki, as well as with Max Haider and Stefan Uhlemann (CEOS GmbH). For the preparation of samples, we are grateful to Kai Rossnagel (1T-TaS$_2$), and Benedikt Iffland, Christian Jooss and Vladimir Roddatis (PCMO). Technical support by Karin Ahlborn and Mathias Hahn is gratefully acknowledged. We thank Felix Börrnert and Hannes Lichte for providing us with the bi-prism. We are also thankful for very productive collaborations with JEOL Ltd., JEOL (Germany) GmbH, and YPS Ltd., and we would like to note the particularly valuable technical support by Bernd Karoske (JEOL (Germany) GmbH).

This work was funded by the Deutsche Forschungsgemeinschaft (DFG) in Collaborative Research Center "Atomic Scale Control of Energy Conversion" (DFG-SFB 1073, project A05) and in the Priority Program "Quantum Dynamics in Tailored Intense Fields" (DFG-SPP1840). We gratefully acknowledge support by the Lower Saxony Ministry of Science and Culture and funding of the instrumentation by the DFG and the VolkswagenStiftung.